# Experimental investigation of the asymmetric spectroscopic characteristics of electron- and hole-doped cuprates

N.-C. Yeh,[a*] C.-T. Chen,[a] A. D. Beyer,[a] C. R. Hughes,[a] T. A. Corcovilos,[a] S. I. Lee[b]

[a]*Department of Physics, California Institute of Technology, Pasadena, CA 91125, USA*

[b]*Department of Physics, Pohang University of Science and Technology, Pohang, Korea*

**Abstract**

Quasiparticle tunneling spectroscopic studies of electron- (n-type) and hole-doped (p-type) cuprates reveal that the pairing symmetry, pseudogap phenomenon and spatial homogeneity of the superconducting order parameter are all non-universal. We compare our studies of p-type $YBa_2Cu_3O_{7-\delta}$ and n-type infinite-layer $Sr_{0.9}Ln_{0.1}CuO_2$ (Ln = La, Gd) systems with results from p-type $Bi_2Sr_2CaCu_2O_x$ and n-type one-layer $Nd_{1.85}Ce_{0.15}CuO_4$ cuprates, and attribute various non-universal behavior to different competing orders in p-type and n-type cuprates.

## 1. Introduction

The presence of competing orders in the ground state of cuprate superconductors [1] results in rich phenomena and complications for unraveling the pairing mechanism. Recent experimental development [1-3] reveals significant non-universal phenomena and asymmetric characteristics between n-type and p-type cuprates as the consequences of competing orders. In particular, the asymmetric characteristics among n-type and p-type cuprates may be attributed to the differences in their low-energy spin excitations [1,2]. We suggest that the *incommensurate* spin excitations associated with charge modulations in *p-type* cuprates may result in a charge nematic (CN) phase that competes with superconductivity (SC), yielding pseudogap phenomena and nano-scale phase separations in two-dimensional (2D) cuprates and long-range SC order in 3D cuprates [1,2]. In contrast, *commensurate* spin excitations in *n-type* cuprates, as manifested by neutron scattering [4] and implied by quasiparticle spectra [2,3], are indicative of the coexistence of antiferromagnetism (AFM) with SC.

## 2. Competing orders and pseudogap in p-type cuprates

The doping of holes into the $CuO_2$ planes of cuprates is known to induce gapped spin excitations in the $CuO_2$ plane. One of the possible ground states of the hole-doped $CuO_2$ plane is the stripe phase which accommodates gapped spin excitations via charge modulations [1]. However, a charge stripe phase with long-range order is energetically very costly. A compromised competing ground state phase in the presence of disorder could be a CN phase that involves local charge modulations while evading strong Coulomb repulsion. In general, the competition between two order parameters can result in three possible phase diagrams as a function of the chemical potential [1-3]: 1) nano-scale phase separations of the two phases, 2) coexistence of the two phases, and 3) disorder intermediate between the two phases. For highly 2D p-type cuprates such as $Bi_2Sr_2CaCu_2O_x$ (Bi-2212), the resulting ground state could reflect either Case 1) or Case 2), because SC involving continuous U(1) symmetry breaking cannot sustain long-range homogeneity in 2D [2], and also because the CN phase could be better stabilized by local disorder in 2D [8].

---

* Corresponding author. Tel.: +1-626-395-4313; fax: +1-626-683-9060; e-mail: ncyeh@caltech.edu. Research at Caltech is supported by NSF.

The CN phase may extend above $T_c$ and yield the pseudogap (PG) phenomenon. In contrast, highly 3D p-type cuprates like YBa$_2$Cu$_3$O$_{7-\delta}$ (YBCO) may belong to Case 2), with homogeneous SC manifested by scanning tunneling spectroscopy [1], NMR [5] and microwave [1] studies. Figure 1 compares the spatial evolution of quasiparticle

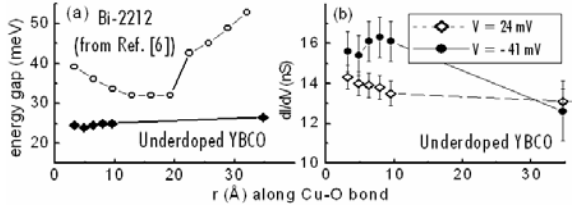

spectral characteristics between YBCO and Bi-2212.

Fig.1 Spatial evolution of quasiparticle spectral features along the Cu-O bonding direction: (a) energy gap of underdoped YBCO ($T_c \approx$ 60 K) [1] and Bi-2212 ($T_c \approx$ 79 K) [6]; (b) differential conductance (dI/dV) of underdoped YBCO at bias voltages V = 24 mV and –41 mV.

To investigate the possibility of locally pinned CN coexisting with SC in Bi-2212, we consider the effects of three sources of elastic scattering on the energy ($E$) and momentum transfer ($q$) dependence of the tunneling conductance $\Delta G(q,E)$, which is proportional to the Fourier-transformed local density of states (FT-LDOS) $\rho_q(E)$ [7,8]: point defects in the SC region, stripes in the pinned CN region, and 1D "edge states" separating the CN and SC regions. Assuming weak scattering potentials and using the first-order T-matrix approximation [8], we find that $\rho_q(E) \propto \int d^3k\ \delta(E-E_k)\ \delta(E-E_{k+q})\ |V(q)|\ F_\pm(k,q)$, where $|V(q)|$ is the elastic scattering matrix in the Born approximation, and $F_\pm(k,q)$ are the coherence factors for spin-independent (+) and spin-dependent (−) interactions [8]. Examples of the quasiparticle FT-LDOS are shown in Figs. 2(a)-(c), where we have included 24 randomly distributed defects in an area of (200$a_0 \times$200$a_0$). We note that the FT-LDOS in Fig. 2(b) due to CN scattering of quasiparticles reveals more intense features along the ($\pi$,0)/(0,$\pi$) directions than those due to point and edge defects in Figs. 2(a) and 2(c). The combined spectra of both point- and 1D-scattering centers seem to agree better with experimental observation [7,8], implying that pinned CN could coexist with SC in Bi-2212.

## 3. Coexisting AFM and SC in n-type cuprates

The presence of commensurate spin excitations in one-layer n-type cuprates, as revealed from neutron scattering data [4], is consistent with the coexistence of a long-range AFM order parameter in the SC state, and the absence of charge modulations may account for the absence of PG above $T_c$ [9]. Further evidence for coexisting AFM and SC is provided by our quasiparticle spectra of the n-type infinite-layer cuprate Sr$_{0.9}$La$_{0.1}$CuO$_2$: For small tunneling currents (< 20 nA), similar spectral characteristics with $\Delta_{SC}$ = 13 meV prevail, whereas for large tunneling currents (> 80 nA), a different large-gap spectrum ($\Delta_{AFM} \sim$ 25 meV) emerges. The field-induced pseudogap below $T_c$ and above the upper critical field $B_{c2}(T)$ for one-layer n-type cuprates [9] is also suggestive of remnant AFM or a spin-flop (SF) phase upon the suppression of SC at $B > B_{c2}(T)$. In Fig. 3 we compare the phase diagram of the n-type infinite-layer (from our magnetization measurements) with that of the one-layer cuprates [9]. The different vortex phase boundaries of Sr$_{0.9}$Gd$_{0.1}$CuO$_2$ and Sr$_{0.9}$La$_{0.1}$CuO$_2$ suggest excess pinning effects due to the magnetic moments of Gd.

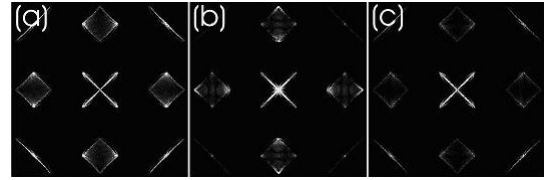

Fig.2 Quasiparticle FT-LDOS associated with (a) non-magnetic point defects, (b) stripes in CN, and (c) edge states for $(q_x,q_y) = (\pm\pi/a_0,\pm\pi/a_0)$, $\Delta_d$ = 40 meV and $E$ = 20 meV. The sharper contrast represents stronger interference intensity.

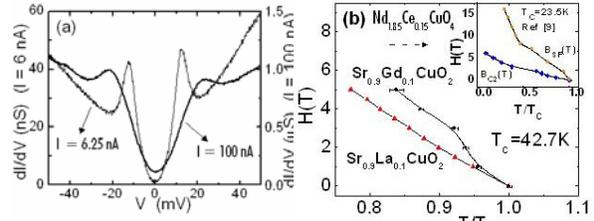

Fig.3 (a) Quasiparticle tunneling spectra of Sr$_{0.9}$La$_{0.1}$CuO$_2$ at 4.2 K taken with tunneling currents of $I$ = 6.25 nA and 100 nA. (b) Magnetic phase diagram of n-type cuprates: infinite-layer (main panel) and one-layer (inset). In the inset $B_{SF}(T)$ refers to a spin-flop phase for H || c-axis.

## References

[1] N.-C. Yeh, Bulletin of Assoc. Asia Pacific Phys. Soc. Vol. 12, No. 2 (2002) 2 [cond-mat/0210656].
[2] N.-C. Yeh and C.-T. Chen, Int. J. Mod. Phys. B 17 (2003) 3575 [cond-mat/0302217].
[3] C.-T. Chen et al., Phys. Rev. Lett. 88 (2002) 227002.
[4] M. Matsuda et al., Phys. Rev. B 66 (2002) 172509.
[5] J. Bobroff et al., Phys. Rev. Lett. 89 (2002) 157002.
[6] K. M. Lang et al. Nature 415 (2002) 412.
[7] J. E. Hoffman et al., Science 297 (2002) 1148.
[8] C.-T. Chen and N.-C. Yeh, Phys. Rev. B 68 (2003) 220505(R) [cond-mat/0307660].
[9] S. Kleefisch et al., Phys. Rev. B 63 (2001) 100507.